\documentclass[12pt,draftcls,onecolumn]{IEEEtran}

\usepackage{cite}

\ifCLASSINFOpdf
   \usepackage[pdftex]{graphicx}
\else
   \usepackage[dvips]{graphicx}
\fi

\usepackage{subfig}
\usepackage{cite}

\begin{document}
\title{Rate-Privacy in Wireless Sensor Networks}

\author{\IEEEauthorblockN{H. Shafiei\IEEEauthorrefmark{1}\IEEEauthorrefmark{2},
A. Khonsari\IEEEauthorrefmark{1}\IEEEauthorrefmark{2},
H. Derakhshi\IEEEauthorrefmark{1},
and 
P. Mousavi\IEEEauthorrefmark{1}} \\
\IEEEauthorblockA{Emails: h.shafiei@ut.ac.ir, ak@ipm.ir, h.derakhshi@ut.ac.ir, pa.mousavi@ut.ac.ir}\\

\IEEEauthorblockA{\IEEEauthorrefmark{1}ECE Department, University of Tehran, Tehran, Iran}\\
\IEEEauthorblockA{\IEEEauthorrefmark{2}School of Computer Science, IPM, Tehran, Iran}}

\maketitle

\begin{abstract}
This paper introduces the concept of rate privacy in the context of wireless sensor networks. Our discussion reveals that the concept indeed  is of a great importance for the privacy preservation of such networks. As a result, we propose a buffering scheme to protect the rate from adversaries. Simulation results verify the applicability of our approach. 

\end{abstract}

\IEEEpeerreviewmaketitle

\newtheorem{MyLemmas}{Lemma}
\newtheorem{MyTheo}{Theorem}

\section{Introduction}
The topic of privacy preservation in Wireless Sensor Networks (WSNs) has attracted many research studies during past few years \cite{li2009privacy}. Two types of privacy have been discussed in the literature; data-oriented privacy and context-oriented privacy. While data-oriented approaches protect the privacy of the content of the sensed data, context-oriented privacy schemes focus on preservation of the context information such as the location of sensors or the time when an event message is generated i.e., temporal privacy \cite{Kamat}.

There are two previous studies that have considered the temporal privacy in WSNs. Kamat et. al., \cite{Kamat} propose a buffering approach maintained at intermediate nodes along the routing path, to inject delay and as a result obfuscate the creation time of each message from the adversary. Using their approach, the adversary cannot extract the exact time when the source node sensed the event which prevents the revelation of many critical information about the target. Hong et. al., \cite{hong2005effective} argue that by detecting the relation between upstream and downstream traffic an adversary can trace the routing path to the region of interest such as the location of base-station. So they propose a random delaying strategy in order to hide the timing correlation between the two streams. Each node randomly delays each received message in a way that it mixes the order of the messages. Both studies consider that the delay follows the exponential distribution.

The above studies protect network's temporal privacy. However, there are some situations in which the answer to the questions like: ``How often a particular event occurs?'' or ``At which rate the event occurs?'' also is a part of the network owners' privacy. For example, consider a factory that has installed a WSN at its production lines. Every time the line finalizes a product an assigned sensor node generates a message to inform the base-station. Realizing the rate at which the factory produces its products is of a great importance for its competitors as they can obtain valuable knowledge e.g., they can predict the market. Another example would be a WSN installed for surveillance purposes at the gates of a garrison. Having the knowledge of the rate at which soliders enter the gates along with some side information results in the ability to determine the number of soliders. Moreover, the enemy can find a time interval during which (with high probability) no one appears at the gate.

Assume that the message flow obeys Poisson distribution with parameter $\lambda$. An adversary can eavesdrop the channel and sample the message transmission rate. She can estimate $\lambda$ using the sample mean. It has been proved that the sample mean of the Poisson distribution is a minimum-variance unbiased estimator as its estimation variance achieves the Cramer-Rao lower bound \cite{papoulis2001probability}.

No other study has elaborated on the temporal privacy of WSN other than the two work mentioned earlier in this paper and neither of these studies can hide the rate from the adversary. In fact, as they consider exponential service time the output is a Poisson process of the same rate (Burke's theorem \cite{papoulis2001probability}).

This paper introduces the concept of rate privacy in WSNs. To the best of our knowledge, no such concept has ever been discussed in the literature. The paper provides a method to protect the actual rate from the adversary.

\begin{figure}[t]
\centering
\includegraphics[width=4in]{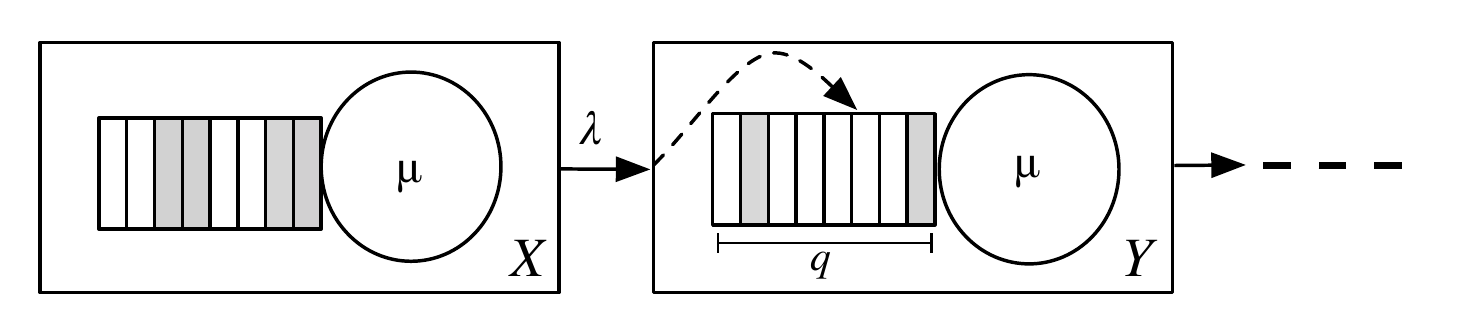}
\caption{An example of sensor nodes in a routing path. Shaded boxes depict occupied slots in each buffer}
\label{fig:q}
\end{figure}

\section{The Proposed Method}
There are two trivial solutions to hide the rate, as follows:

\begin{enumerate}
\item  Deterministic message generation: Each node constantly senses the area, however, it generates the corresponding messages only in deterministic time intervals e.g., every 10 minutes. This approach is not adapted to the sensing events. Large intervals delay the message while small delay intervals burden worthless traffic to the network.
\item Randomized dummy messages: This approach increases the energy consumption as every node periodically generates dummy messages.
\end{enumerate}

\begin{figure*}[t]
\subfloat[Buffer length versus the estimated rate]{\includegraphics[width=2.92in]{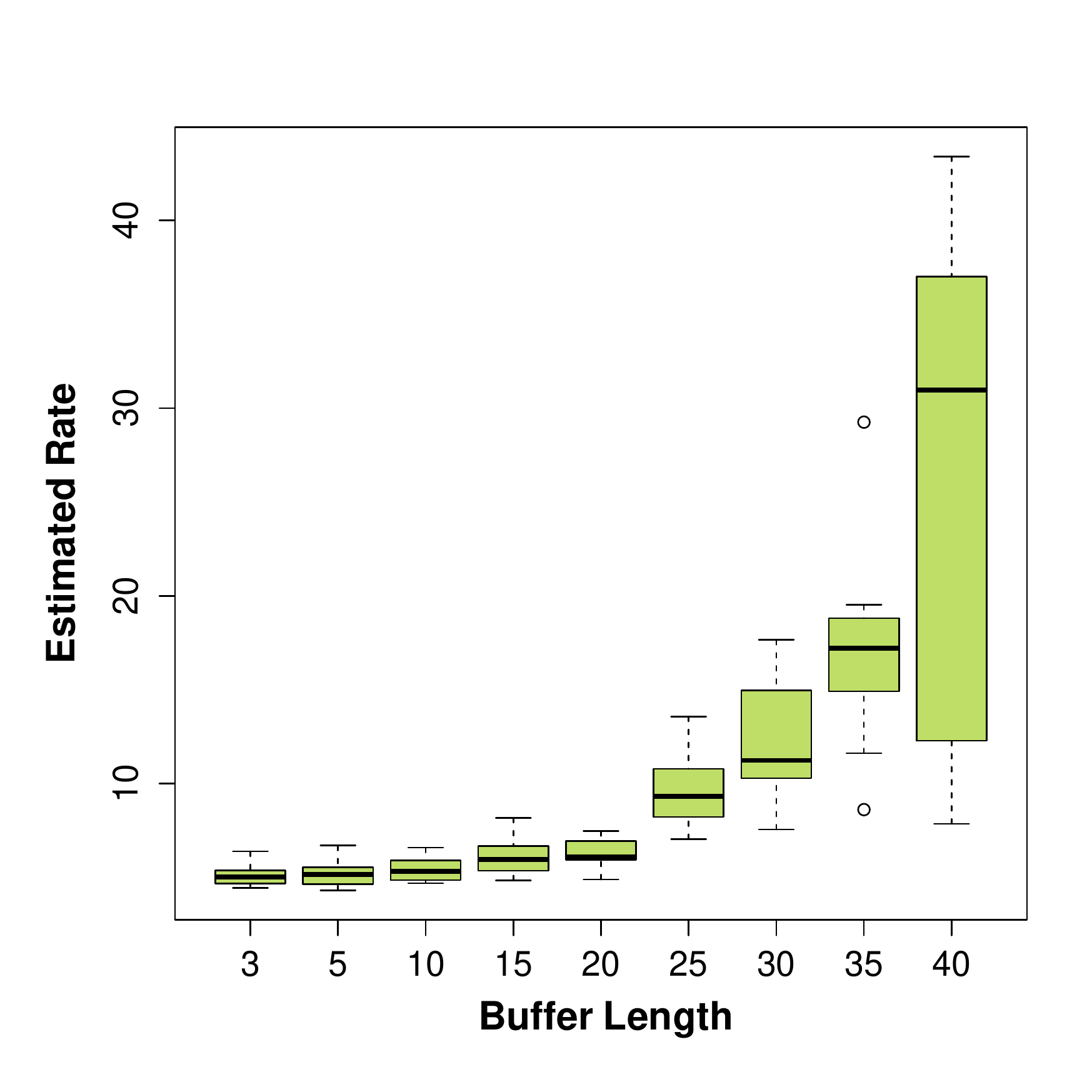}}
\hfill
\subfloat[Buffer length versus the average message latency]{\includegraphics[width=2.92in]{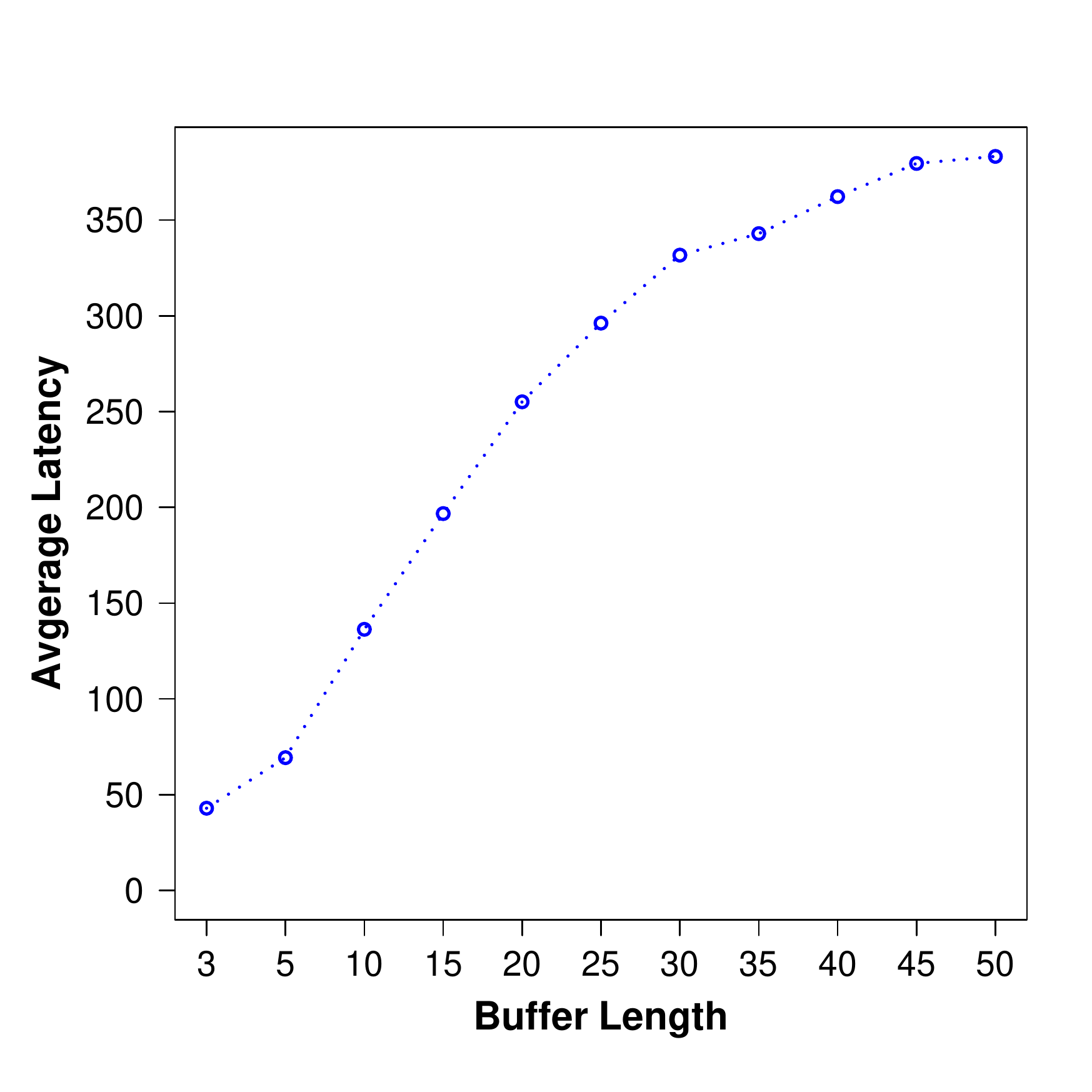}}

\subfloat[Simulation time versus the estimated rate]{\includegraphics[width=2.92in]{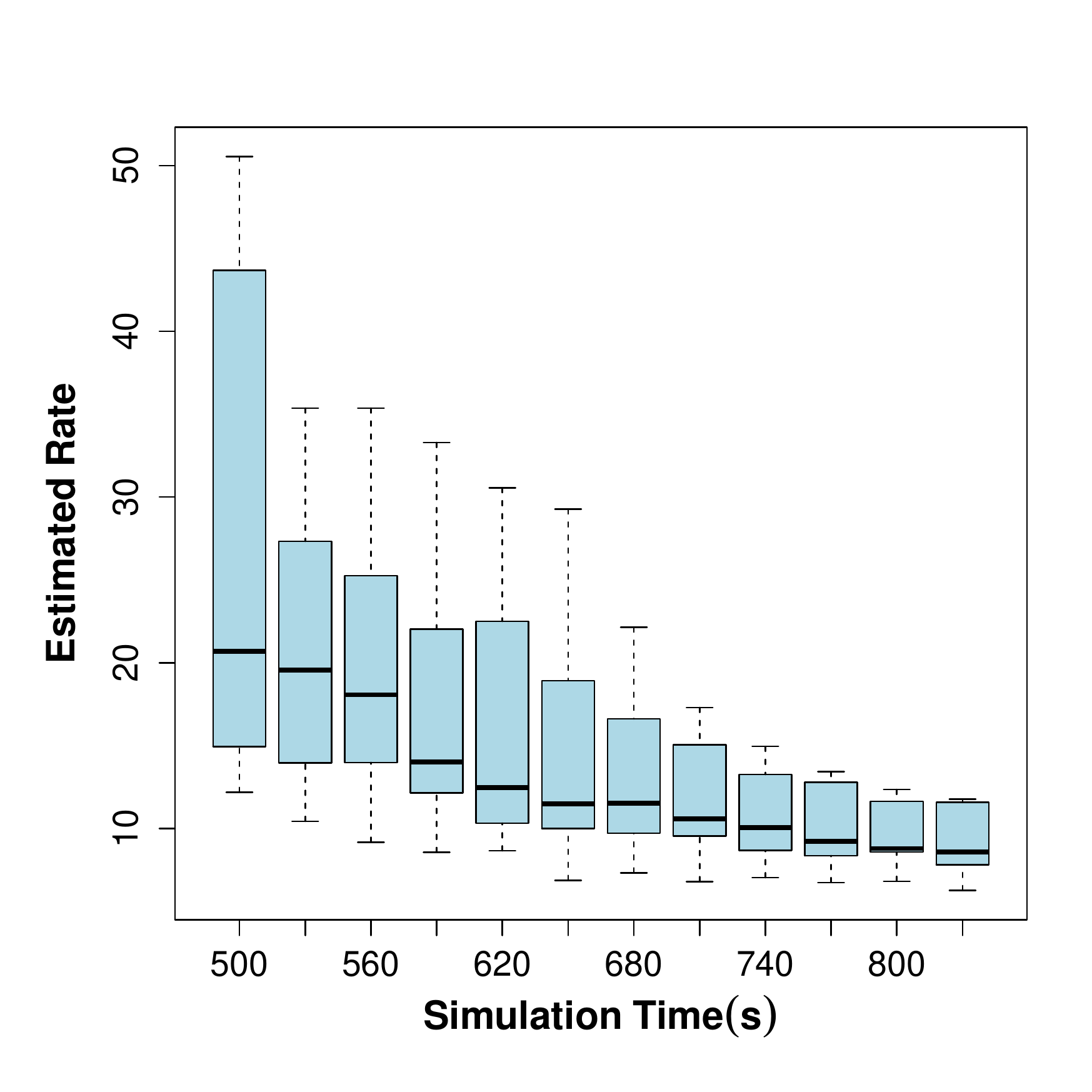}}
\hfill
\subfloat[Inter-arrival time versus average estimation error]{\includegraphics[width=2.92in]{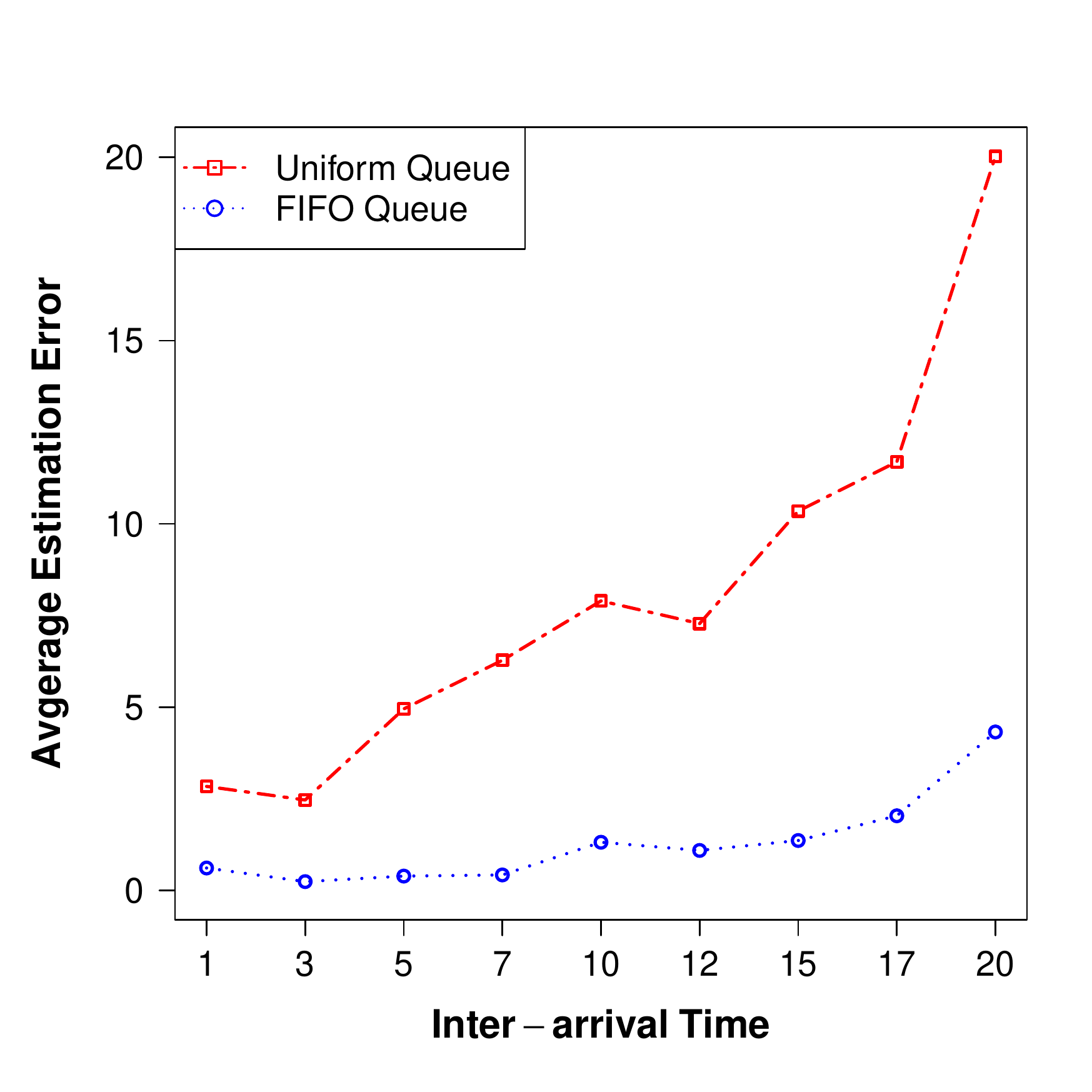}}
\hfill
\caption{Simulation results}
\label{subfig:simulation}
\end{figure*} 

Motivated by the shortcomings of the approaches described above, we present a rate-privacy preservation scheme as follows. Consider that the network is secured and an adversary neither has intention nor has the ability to decrypt the messages. However, the adversary stays near the base-station and eavesdrops the traffic originated from the region in order to estimate the rate without decrypting the messages. All of the sensor nodes independently generate sensing reports using constant size packets, and route these packets using hop-by-hop routing fashion toward the base-station.

In our approach we assume that every node maintains a buffer of size $q$ and each message is buffered in intermediate nodes along the routing path from a source node to the base-station. We change the buffering scheme from First-In-First-Out (FIFO) to a randomized approach using a probability distribution such as uniform distribution. In this scheme, upon reception of a message instead of placing it at the end of the queue, the receiver places the message in one of the empty slots of the buffer uniformly at random. Each node retrieves the first message in the buffer after a random delay that follows the exponential distribution with ($\mu$) and forwards it toward the base-station. Fig. \ref{fig:q} shows an example of such scheme. Node $X$ senses the area, generates report messages with rate $\lambda$ and routes them toward the base-station through its neighbor which is the node $Y$. Upon reception, $Y$ places the received message in one of the empty slots with probability $1/(q - i)$ where $i$ is equal to the number of occupied slots. This approach not only preserves the temporal privacy of the network (as it delays every message) but also protects the rate from the adversary. We leave the modeling of this approach to our future work.

%\section{Analytical Model}

%\begin{MyTheo}
%\label{teo:nmr}
%Let $E[\Delta u_{i-1,i}$ be the average departure time of two consecutive packets from a queue. The attacker can estimate the rate $\lambda$ as follows:
%
%\begin{equation}
%\lambda^{-1} = E[\Delta u_{i-1,i}] - \int _{0}^{\infty} e^{-\lambda \vartheta}\left(1-\mathit{L}^{-1}\left\{ \frac{s(1-\rho)}{s-\lambda +\lambda B^{*}(s)} \right\}_{\vartheta}\right)\mathrm{d}\vartheta
%\end{equation}
%
%\begin{proof}
%\end{proof}
%\end{MyTheo}

\section{Simulation Results}
Our proposed approach has been implemented in Castalia simulator \cite{castalia}. Castalia is a discrete event simulator for sensor networks based on OMENT++ \cite{varga2001omnet++}. Numerous validation experiments have been established. However, for the sake of specific illustration, validation results are presented for limited number of scenarios. We assumed collision-free packet transmission where $100$ nodes are scattered in a $100 \times 100$ $m^2$ area. Moreover, we assumed that the adversary stays near the base-station and observes every received message.

Fig. \ref{subfig:simulation}(a) shows the effect of buffer size on the adversary's estimated rate when the inter-arrival time equals to 5 seconds. The wider interquartile range of the larger buffers confirms that the probability that the adversary correctly estimates the rate reduces significantly as the buffer size grows. Fig. \ref{subfig:simulation}(b) shows the average message latency when buffer length increases.

Fig. \ref{subfig:simulation}(c) depicts the snapshot of the last 300 seconds of the simulation when the inter-arrival time is 5 and the buffer size is 20. We assumed that the adversary samples the inter-arrival times during the entire simulation time. It can be observed that as the sample size grows the estimation becomes more accurate, however, the size of the first quartile suggests that with high probability the estimated rate at most reaches to a value which is 2 times greater than the actual rate.

Fig. \ref{subfig:simulation}(d) compares the average estimation error when the inter-arrival rate varies, for two different cases; our proposed scheme and a FIFO-based buffering approach \cite{Kamat}. It can be realized that in the latter case the estimation error is negligible. However, the average error in our approach is significantly greater which prevents the adversary from successfully estimating the actual rate. Also note that the error increases when the  inter-arrival rate raises which is mainly due to the smaller sample size of uncrowded networks.
\section{Conclusion}

In this paper the concept of rate-privacy in wireless sensor networks is introduced. We presented a buffering approach that preserves both rate and temporal privacy. Our next steps target to elaborate on the mathematical modeling of the proposed approach. Also, rate-privacy protection in realtime sensor networks is a good direction for future studies.
\bibliographystyle{IEEEtran}
\bibliography{IEEEabrv,shafiei}

\end{document}